\documentclass[12pt]{article}
\begin{document}

\title{{\bf Agnesi Weighting for the 
 \\Measure Problem of Cosmology}
\thanks{Alberta-Thy-13-10, arXiv:1011.4932 [hep-th]}}

\author{
Don N. Page
\thanks{Internet address:
profdonpage@gmail.com}
\\
Theoretical Physics Institute\\
Department of Physics, University of Alberta\\
Room 238 CEB, 11322 -- 89 Avenue\\
Edmonton, Alberta, Canada T6G 2G7
}

\date{2011 February 24}

\maketitle
\large
\begin{abstract}
\baselineskip 25 pt

The measure problem of cosmology is how to assign normalized
probabilities to observations in a universe so large that it may have
many observations occurring at many different spacetime locations.  I
have previously shown how the Boltzmann brain problem (that observations
arising from thermal or quantum fluctuations may dominate over ordinary
observations if the universe expands sufficiently and/or lasts long
enough) may be ameliorated by volume averaging, but that still leaves
problems if the universe lasts too long.  Here a solution is proposed
for that residual problem by a simple weighting factor $1/(1+t^2)$ to
make the time integral convergent.  The resulting Agnesi measure appears
to avoid problems other measures may have with vacua of zero or negative
cosmological constant.  

\end{abstract}

\normalsize

\baselineskip 17.6 pt

\newpage

\section*{Introduction}

The high degree of spatial flatness observed for the constant-time
hypersurfaces of our universe leads to the idea that our universe is
much larger than what we can presently observe.  The leading explanation
for this flatness, cosmological inflation in the early universe,
suggests that in fact the universe is enormously larger than what we can
see, perhaps arbitrarily large if an indefinitely long period of eternal
inflation has occurred in the past.  Furthermore, the recent
observations of the acceleration of the universe suggest that our
universe may expand exponentially yet more into a very distant future. 
As a result, spacetime may already be, or may become, so large that a
vast number of different observations (by which I mean observational
results, what it is that is actually observed) will recur a huge number
of times throughout the universe. 

If we restrict to theories that only predict whether or not a particular
observation (e.g., ours) occurs, there are likely to be many such
theories predicting that our observation almost certainly occurs, so
that we would have very little observational evidence to distinguish
between such theories.  This would seem to imply that observational
science would come to an end for such theories.

However, even for a very large universe there can be theories that are
much more testable by predicting not just whether a particular
observation occurs, but also the probability that this particular
observation is made rather than any of the other possible observations. 
Then one can use the probability the theory assigns to our actual
observation as the likelihood of the theory, given the observation
(actually the conditional probability of the observation, given the
theory).  One can then draw statistical conclusions about alternative
theories from their likelihoods.  For example, in a Bayesian analysis in
which one assigns prior probabilities to the theories, one can multiply
these priors by the likelihoods and divide by the sum of the products
over all theories to get the normalized posterior probabilities of the
theories.

Therefore, it would be desirable to have theories that each predict
normalized probabilities for all possible observations.  (These
probabilities can be normalized measures for a set of observations that
in a global sense all actually occur, as in Everettian versions of
quantum theory in which quantum probabilities are not propensities for a
wide class of potentialities to be converted to a narrower class of
actualities.  All observations with positive measure could actually
occur in such a completely deterministic version of quantum theory, but
with different measures, which, if normalized, can be used as
likelihoods in a statistical analysis.)

However, in a very large universe in which many observations recur many
times, it can become problematic what rule to use to calculate the
normalized measure (or probability) for each one.  If one had a definite
classical universe in which each observation occurs a fixed finite
number of times, and if the total number of all observations is also a
finite number, one simple rule would be to take the normalized measure
for each observation to be the fraction of its occurrence, the number of
times it occurs divided by the total number of all observations.  But in
a quantum universe in which there are amplitudes for various situations,
it is less obvious what to do.

I have shown that Born's rule, taking the normalized measure of each
observation to be the expectation value of a corresponding projection
operator, does not work in a sufficiently large universe
\cite{Born1,Born2,Born3,Born4,Born5}.  The simplest class of replacement
rules would seem to be to use instead the expectation values of other
positive operators, but then the question arises as to what these
operators are.

For a universe that is a quantum superposition of eigenstates that each
have definite finite numbers of each observation, one simple choice for
the normalized measures would be to take the expectation values of the
frequencies of each observation (say frequency averaging), and a
different simple choice would be to take the expected numbers of each
observation divided by the expected total number of all observations
(say number averaging).  For example, suppose that the quantum state
giving only two possible observations (say of a loon or of a bear, to
use the animals on the one- and two-dollar Canadian coins) is
\begin{equation} 
|\psi\rangle = \cos{\theta}|mn\rangle +\sin{\theta}|MN\rangle,  
\label{2-state} 
\end{equation} 
where the first eigenstate $|mn\rangle$ corresponds to $m$ loon
observations and $n$ bear observations and the second eigenstate
$|MN\rangle$ corresponds to $M$ loon observations and $N$ bear
observations.  (For simplicity I am assuming all of the loon
observations are precisely identical but different from all of the bear
observations that are themselves precisely identical.)  Then the first
choice above, frequency averaging, would give 
\begin{eqnarray}
P_f(\mathrm{loon})&=&\frac{m}{m+n}\cos^2{\theta}
                    +\frac{M}{M+N}\sin^2{\theta}, \nonumber \\ 
P_f(\mathrm{bear})&=&\frac{n}{m+n}\cos^2{\theta}
                    +\frac{N}{M+N}\sin^2{\theta}, 
\label{frequency}
\end{eqnarray} 
whereas the second choice above, number averaging, would give 
\begin{eqnarray}
P_n(\mathrm{loon})&=&\frac{m\cos^2{\theta}+M\sin^2{\theta}}
{(m+n)\cos^2{\theta}+(M+N)\sin^2{\theta}}, \nonumber \\ 
P_n(\mathrm{bear})&=&\frac{n\cos^2{\theta}+N\sin^2{\theta}}
{(m+n)\cos^2{\theta}+(M+N)\sin^2{\theta}}. 
\label{number}
\end{eqnarray}
Therefore, even in this very simple case, there is no uniquely-preferred
rule for converting from the quantum state to the observational
probabilities.  One would want $P(\mathrm{loon})$ to be between the two
loon-observation frequencies for the two eigenstates, between $m/(m+n)$
and $M/(M+N)$, as indeed both rules above give, but unless one believes
in the collapse of the wavefunction (which would tend to favor frequency
averaging), there does not seem to be any clear choice between the two. 
(One can easily see that in this example, there is no state-independent
projection operator whose expectation value is always between $m/(m+n)$
and $M/(M+N)$ for arbitrary $m$, $n$, $M$, $N$ and $\theta$, so Born's
rule fails \cite{Born1,Born2,Born3,Born4,Born5}.)

The problem becomes even more difficult when each quantum component may
have an infinite number of observations.  Then it may not be clear how
to get definite values for the frequencies of the observations in each
eigenstate, or how to get definite values for the ratios of the infinite
expectation values for the numbers of each different observation.  Most
of the work on the measure problem in cosmology has focused on
regularizing these infinite numbers of observations
\cite{Linde86,LinMez,BLL94,Vil95a,LLM,GL,LLM2,Vil95b,WV,LM,Vil98,VVW,
Guth00,GV,EKS,SEK, AT,Teg,Aguirre,Ellis,GSVW,ELM,BFL,Bou06,CDGKL,BFY06,
GT,Vil07,AGJ1,Win06,AGJ2,BHKP,Guth07,BY,LWa,ADNTV,Linde08a,Linde08b,LWb,
CSS,Haw,GV07,HHH1,HHH2,HHH3,HHH4,Win08a,SGSV,Win08b,Born1,BFYb,DGLNSV,
GV09,Win08,LVW,Bou09,Sal09,BY09,SH1,SH2,Pagedecay,SeSu,LV10,SiSa,
Gratton,SPV,BFLR1,LN,NV,BFLR2,LV,BFLR3,BFLR4}.
However, I have emphasized \cite{Born1,Born2,Born3,Born4,Born5}
that there is also the ambiguity described above even for finite numbers
of occurrence of identical observations.

One challenge is that many simple ways to extract observational
probabilities from the quantum state appear to make them dominated by
Boltzmann brain observations, observations produced by thermal or vacuum
fluctuations rather than by a long-lived observer
\cite{DKS,Albrecht,AS,Page05,YY,Page06a,BF,Page06b,Linde06,Page06c,
Vil06,Page06d,Vanchurin,Banks,Carlip,HS,GM,Giddings,typdef,LWc,DP07,
Bou08,BFYa,ADSV,Gott,typder,FL,NYTimes,DO}. But if Boltzmann brains
dominate, we would expect that our observations would have much less
order than they are observed to have, so we have strong statistical
evidence against our observations' being made by Boltzmann brains.  We
would therefore like theories that does not have the measures for
observations dominated by Boltzmann brains.

The main way in which theories tend to predict domination by Boltzmann
brains is by having the universe last so long that after ordinary
observers die out, a much larger number (say per comoving volume) of
Boltzmann brains eventually appear.  A big part of the problem is that
the volume of space seems to be beginning to grow exponentially as the
universe enters into an era in which a cosmological constant (or
something very much like it) dominates the dynamics.  Therefore, the
expected number of Boltzmann brain observations per unit time would grow
exponentially with the expansion of the universe and would eventually
become larger than the current number of observations per time by
ordinary observers, leading to Boltzmann brain domination in number
averaging (which I have previously called volume weighting
\cite{Born1,Born2,Born3,Born4,Born5} because the number per time is
proportional to the spatial volume for observations at a fixed rate per
four-volume).

To avoid this part of the problem that occurs for what I have called
volume weighting (or what I now prefer to call number averaging, setting
the measure for a observation proportional to the expectation value or
quantum average of the number of occurrences of the observation), I have
proposed using instead volume averaging
\cite{Born1,Born2,Born3,Born4,Born5} (or what I now prefer to call
spatial density averaging), setting the measure for each observation on
a particular hypersurface to be proportional to the expectation value of
the spatial density of the occurrences of that observation, the expected
number divided by the spatial volume.  This would lead to the
contribution per time for hypersurfaces at late times being the very low
asymptotically constant spacetime density of Boltzmann brains. This
density is presumably enormously lower than the spacetime density of
ordinary observers today, so per time, observations today dominate
greatly over Boltzmann brains at any time in the distant future.

However, if the universe lasts for a sufficiently long time
(exponentially longer than the time it would have to last for Boltzmann
brains to dominate in number averaging), then integrating over time
would cause even the contribution from the very tiny spatial density of
Boltzmann brains eventually to dominate over the contributions of
ordinary observers that presumably exist only during a finite time. 
(For the moment I am ignoring the contributions from tunnelings to new
vacua, which will be discussed below.)  Therefore, it appears that we
need not only a shift from number averaging (volume weighting) to
spatial density averaging (volume averaging), but that we also need
something else to suppress the divergence in the Boltzmann brain
contributions at infinite times.

In the scale-factor measure \cite{SGSV,BFYb,DGLNSV,SiSa}, one puts a
cutoff where the volume of cross sections of certain sets of timelike
geodesics reaches some upper limit.  This avoids the divergence one
would get from number averaging (volume weighting) without a cutoff, in
which the volume and the number of observations would go to infinity as
the time is taken to infinity.  By putting a cutoff on the volume, the
scale-factor measure gives results rather similar to spatial density
averaging (volume averaging), because for exponentially expanding
universes, most of the observations occur near the cutoff where the
volume is fixed, so that the total number of observations is essentially
proportional to the spatial density of observations (up to a factor for
the amount of time over which most of the observations occur before the
cutoff, which is inversely proportional to the Hubble expansion rate if
it is constant).

If the universe is dominated by a positive cosmological constant at late
times, geodesics other than those that stay within bound matter
configurations expand indefinitely at an asymptotically exponential
rate, so that all such geodesics eventually reach the cutoff.  Then if
the contributions within the bound matter configurations, where the
geodesics are not cut off, do not dominate, then one only gets a finite
set of observations of each type, and one can apply either frequency
averaging or number averaging.  (One might expect the matter in bound
matter configurations eventually to decay away, so that one does not
have to worry about the timelike geodesics there that would never expand
to the cutoff if the matter configuration persisted.)

The usual answer that one gets is that if the universe tends to a
quasi-stationary eternal inflation picture in which new bubbles are
forming and decaying at an asymptotically fixed rate, and if the cutoff
is applied at a sufficiently great volume, then the precise value at
which it is applied does not matter \cite{SGSV,BFYb,DGLNSV,SiSa}. 
Furthermore, if Boltzmann brains do not form at all in the longest-lived
de Sitter vacuum, and if the tunneling rate to new bubbles that lead to
more ordinary observers is greater than the rate for Boltzmann brains to
form in {\it all} anthropic vacua, then ordinary observers dominate over
Boltzmann brains \cite{BFYb,DGLNSV}.  (See analogous restrictions
\cite{Linde06,BF} for other measures that unfortunately I do not have
time to discuss in detail in this paper.)  Although it is not
implausible that this latter requirement may be satisfied in a large
fraction of anthropic vacua \cite{FL}, it does seem quite stringent for
it to be satisfied in {\it all} of them if there are an exponentially
large number. Therefore, the scale-factor measure still has a potential
Boltzmann brain problem.

Another somewhat undesirable feature of the scale-factor measure (at
least from my viewpoint, though others like Bousso disagree
\cite{Bousso}) is that it is important in this approach that
there be a cutoff, which is crucial for defining the ensemble of
observations.  It has even been noted \cite{BFLR2} that using this
cutoff, ``Eternal inflation predicts that time will end.''

In the scale-factor measure, it is not specified precisely where the
cutoff is to be imposed (at what volume, relative to some initial volume
of each set of timelike geodesics), but it is just pointed out that the
resulting observational probabilities appear to be insensitive to the
value of the cutoff so long as it is sufficiently late (or large). 
However, for a precise theory with a cutoff, one might like a precise
cutoff (though the procedure of taking the cutoff to infinity does
appear to give precise statistical predictions).  It then seems to me a
bit {\it ad hoc} to have the cutoff at some particular very late (or
large) value, as seems to be necessary with the scale-factor cutoff,
unless one can really understand what it means for the cutoff to be
taken to infinity.  (What simple explanation could be given for the very
large value of the cutoff time if it is finite, and on the other hand,
what can be meant by a cutoff if it is at infinite time?)

Here I am proposing to replace the scale-factor cutoff that has an
unspecified or infinite late value with a particular simple explicit
weighting factor to suppress the measures for late-time observations,
such as Boltzmann brains, in a precisely specified way.  The idea is to
supplement the spatial density averaging (volume averaging), which
greatly ameliorates the Boltzmann brain problem, with a measure over
time that integrates to a finite value over infinite time.  The measure
over time is chosen to be $dt/(1+t^2)$, where $t$ is the proper time in
Planck units, which is the simplest analytic weighting I could think of
that gives a convergent integral over time.  Since the curve $y =
1/(1+x^2)$ is named the witch of Agnesi, I shall call this Agnesi
weighting.  (The witch of Agnesi was named after the Italian linguist,
mathematician, and philosopher Maria Gaetana Agnesi, 1718-1799, after a
misidentification of the Italian word for ``curve'' with the word for
``woman contrary to God,'' so that it was mistranslated ``witch.'') 

With an appropriate quantum state, this Agnesi weighting appears to be
consistent with all observations and in particular avoids the potential
Boltzmann brain problem remaining with the scale-factor cutoff if not
all anthropic vacua give tunneling rates (eventually leading to new
anthropic bubbles and new ordinary observers) greater than the rate of
Boltzmann brain formation \cite{Linde06,BF,BFYb,DGLNSV,FL}.  It also
appears to avoid problems that other measures have with zero or negative
cosmological constant \cite{BFLR4}, as shall be discussed below.

\baselineskip 17.2 pt

\section*{Probabilities of observations with Agnesi weighting}

In this paper I shall use a semiclassical approximation for gravity,
since I do not know how to do Agnesi weighting in full quantum gravity. 
Assume that the quantum state corresponds to a superposition of
semiclassical spacetime geometries.  Further assume that the postulated
operators whose expectation values give the measures for observations
commute approximately with the projection operators to the semiclassical
geometries, so that for the measures one can regard the quantum state as
if it were an ensemble of 4-geometries with probabilities $p(^4g)$ given
by the absolute squares of the amplitudes for each geometry.  There is
no guarantee that this approximation is good, but here I shall make it
for simplicity.  Perhaps later one can go back and look at refinements,
though it may be hard to do that without knowing more about the
postulated operators.

I shall assume that each semiclassical 4-geometry has a preferred
fiducial Cauchy hypersurface.  In a standard big-bang model, this could
be the singular surface at the big bang.  In my Symmetric-Bounce model
for the universe \cite{Symbounce}, which I have argued is more
predictive, the fiducial hypersurface would be the hypersurface in which
the semiclassical geometry has zero trace of the extrinsic curvature. 
(In this model, to semiclassical accuracy, the entire extrinsic
curvature would vanish on this hypersurface of time symmetry.)  If one
had a different semiclassical model in which there is a bounce rather
than a singular big bang, the fiducial hypersurface could be the
hypersurface of zero trace of the extrinsic curvature, the one that
minimizes the spatial volume if there are more than one such extremal
hypersurfaces.  It shall be left to the future to extend Agnesi
weighting to semiclassical spacetimes that do not have such preferred
fiducial Cauchy hypersurfaces, but one might assume that the quantum
state of the universe, if indeed it leads to semiclassical spacetimes at
all, would lead to semiclassical spacetimes having such preferred
fiducial Cauchy hypersurfaces.  Or, one could just restrict attention to
such quantum states.

Then for each point of the spacetime, I shall choose the simplest choice
of a time function $t$, the proper time of the longest timelike curve
from that point to the fiducial hypersurface.  This will be a timelike
geodesic intersecting the hypersurface perpendicularly.  If there are
two sides to the hypersurface, as in my Symmetric-Bounce model,
arbitrarily take $t$ positive on one side of it (its future) and
negative on the other side of it (its past).  Take the preferred
foliation of the spacetime given by the hypersurfaces of constant $t$. 

These foliation hypersurfaces may have kinks where one goes from one
region of the spacetime with one smooth congruence of timelike geodesics
that maximize the proper time to the fiducial hypersurface to another
region with a discontinuously different smooth congruence of geodesics,
but they are spatial hypersurfaces, not only locally but also globally
in the sense that they are acausal, with no points on them being null or
timelike separated.  It is easy to see that they are semi-spacelike or
achronal (no points timelike separated \cite{HE}), since if any point
$p$ on a foliation hypersurface (say to the future of the fiducial
hypersurface; replacing ``future'' with ``past'' everywhere gives the
same argument for the opposite case) were to the future of any other
point $q$ on the foliation hypersurface by a positive proper time
$\tau$, a timelike curve from $p$ back to the fiducial hypersurface
could go through $q$ and hence be longer by $\tau$ than the longest
timelike curve from $q$, in contradiction to the assumption that each
point on the hypersurface has the same maximal proper time $t$ back to
the fiducial hypersurface.  With a bit more work, one can also get a
contradiction if any point $p$ on the foliation hypersurface were to the
null future of any other point $q$ on the foliation hypersurface (so
that there exists a future-directed null curve from $q$ to $p$, but no
timelike curve), since one could perturb the null curve going back from
$p$ to $q$ to a timelike curve going back from $p$ to a point $r$ on the
longest timelike curve from $q$ back to the fiducial hypersurface, and
this timelike curve can be chosen to have proper time from $p$ back to
$r$ longer than the longest timelike curve from $q$ back to $r$.

Let $V(t)$ be the spatial 3-volume of each such foliation hypersurface,
at a maximal proper time $t$ to the future or to the past of the
fiducial Cauchy surface.  I shall assume that $V(0)$ at $t=0$ is a local
minimum of the spatial volume, the fiducial hypersurface which can have
$V(0) > 0$ in a bounce model or $V(0) = 0$ in a big bang model.  The
proper 4-volume between infinitesimally nearby hypersurfaces of the
foliation is $dV_4 = V(t)dt$.

In a WKB approximation to the Wheeler-DeWitt equation for canonical
quantum gravity, the absolute square of the wavefunctional for the
hypersurfaces integrated over an infinitesimal sequence of hypersurfaces
in a foliation is proportional to the conserved WKB flux multiplied by
the infinitesimal proper time between hypersurfaces \cite{HawPageWKB},
so here I shall take the quantum probability for the hypersurface to be
one of the foliation hypersurfaces between $t$ and $t+dt$ to be
$p(^4g)dt$.  Note that for semiclassical 4-geometries that have $t$
running to infinity, the integral of the absolute square of the
wavefunctional diverges when integrated over the hypersurfaces
corresponding to all $t$.  This fact also suggests the need to put in a
weighting factor or do something else to get finite observational
probabilities out from a quantum state of canonical quantum gravity.

Let us assume that the semiclassical spacetime $^4g$ gives a spacetime
density expectation value $n_j(t,x^i)$ for the observation $O_j$ to
occur at the time $t$ and spatial location $x^i$.  Let $\bar{n}_j(t)$ be
the spatial average of $n_j(t,x^i)$ over the spatial hypersurface.  Then
the expected number of occurrences of the observation $O_j$ between $t$
and $t+dt$ is $dN_j = \bar{n}_j(t)V(t)dt$.  If we were doing number
averaging (volume weighting), we would seek to integrate $dN_j$ over $t$
to get a measure for the observation $O_j$ contributed by the
semiclassical geometry $^4g$ if it were the only 4-geometry.  However,
if $t$ can go to infinity, this integral would diverge.  If $V(t)$ grows
exponentially with $t$, it would still diverge even if we included the
weighting factor $1/(1+t^2)$.  One would need an exponentially
decreasing weight factor (with a coefficient of $t$ in the exponential
that is greater than the Hubble constant of the fastest expanding vacuum
in the landscape) to give a convergent integral if one just used a
function of $t$ with number averaging.  Such a rapidly decaying weight
factor would lead to the youngness problem \cite{Guth07}.

Things are much better if we use spatial density averaging (volume
averaging), which divides $dN_j$ by $V(t)$ to get $\bar{n}_j(t) dt$, the
spatial average of the density of the observation $O_j$ multiplied by
the proper time $dt$.  If we then combine this spatial density averaging
over the spatial hypersurfaces with Agnesi weighting for the time, we
get that the semiclassical 4-geometry $^4g$ contributes $\int
\bar{n}_j(t) dt/(1+t^2)$ to the measure for $O_j$.  Next, we sum this
over the quantum probabilities of the 4-geometries $^4g$ to get the
relative probability of the observation $O_j$ as
\begin{equation} 
p_j = \sum_{^4g} p(^4g)\int \bar{n}_j(t) \frac{dt}{1+t^2}
    = \sum_{^4g}p(^4g)\int\frac{dN_j}{dt}\frac{1}{V(t)}\frac{dt}{1+t^2}.
\label{Agnesi} 
\end{equation} 
Here, of course, the expectation value of the spatially averaged density
$\bar{n}_j(t)$ of the observations $O_j$, and thus also the expectation
value of the rate of observations per time $dN_j/dt$, depend implicitly
on the 4-geometry $^4g$, and by a semiclassical 4-geometry I am
including the quantum state of the matter fields on that 4-geometry, on
which the expectation value $n_j(t,x^i)$ and hence $\bar{n}_j(t)$ and
$dN_j/dt$ are likely to depend, as well as on the 4-geometry itself.

Finally, we get the normalized probabilities for the observations $O_j$
by dividing by the sum of the unnormalized relative probabilities $p_j$:
\begin{equation} 
P_j = \frac{p_j}{\sum_k p_k}.
\label{probability} 
\end{equation} 

It may be noted that the weighting factor $1/[V(t)(1+t^2)]$ in the last
expression of Eq.\ (\ref{Agnesi}) from spatial density averaging and
Agnesi weighting is essentially a nonuniform xerographic distribution in
the language of Srednicki and Hartle \cite{SH1,SH2}.

\section*{Consequences of Agnesi weighting for our universe}

The combination of spatial density averaging (volume averaging) and
Agnesi weighting (so that the expectation value $dN_j/dt$ of the number
of observations $O_j$ per proper time is divided by both the 3-volume
$V(t)$ of the hypersurface and by the Agnesi factor $1+t^2$) avoids all
divergences (assuming that there is a finite upper bound on the spatial
density of observations, as seems highly plausible).  The results appear
to be better than several other recent measures, such as the
scale-factor measure \cite{SGSV,BFYb,DGLNSV,SiSa}, the closely related
fat-geodesic measure \cite{BFYb}, the causal-patch measure
\cite{Bou06,Bou09,BFLR1}, the stationary measure \cite{Linde08b,LVW},
and the apparent-horizon measure \cite{BFLR4}.

In particular, Agnesi weighting solves the Boltzmann brain problem
without having to assume that all anthropic vacua give tunneling rates
to new vacua greater than the rate of Boltzmann brain formation
\cite{Linde06,BF,BFYb,DGLNSV,FL}.  The spatial density averaging avoids
the domination of Boltzmann brains on individual hypersurfaces, no
matter how large, and the Agnesi factor suppresses the the cumulative
contributions of the arbitrarily many hypersurfaces that occur at very
late times.

Agnesi weighting also avoids the potential problem with possible
Boltzmann brains in a supersymmetric 11-demensional Minkowski vacuum. 
Initially, Boltzmann brains were postulated to arise by thermal
fluctuations in asymptotically de Sitter spacetime
\cite{DKS,Albrecht,AS}, but then I pointed out that if observations are
given by the expectation values of localized operators (e.g., a weighted
sum of localized projection operators to have a particular brain
configuration in a finite region of space), Boltzmann brain observations
should also occur with positive probability even in the vacuum
\cite{Page05,Page06b,Page06c,Page06d}.  This would particularly appear
to pose a problem if the states in the landscape can tunnel to a
supersymmetric 11-demensional Minkowski vacuum, since it would have an
infinite volume in which Boltzmann brains might form.  (Even if one
thought that the fields corresponding to the normal excitations of this
vacuum were unable to support observations, surely there would be some
positive expectation values for a finite region to have the right
fluctuations of whatever fields are necessary to give observations
\cite{Hartle}.)

The other measures mentioned above do not seem to suppress the
contributions of such Boltzmann brains.  For example, the scale-factor
measure could have the congruence of geodesics entering into the
Minkowski vacuum with arbitrarily small divergence, in which case
without the repulsive effects of a cosmological constant, they can go
arbitrarily far into the Minkowski spacetime, and sample an arbitrarily
large 11-volume, before they reach the scale-factor cutoff.  Similarly,
the causal-patch measure could be dominated by Boltzmann brains in an
arbitrarily large causal patch corresponding to a geodesic that lasts
infinitely long in the Minkowski vacuum.

Of course, it is not absolutely certain that Boltzmann brains do form in
the vacuum \cite{DO}, and recently Edward Witten told me \cite{Witten}
he did not believe that Boltzmann brains form in a vacuum, where
information processing and dissipation do not occur.  Despite this
expert opinion, it is still hard for me to be convinced that localized
observations would not occur by purely vacuum fluctuations.  If they
can, it is encouraging that Agnesi weighting would explain why they do
not dominate, even if other measures do not.

The other measures mentioned above also appear to have problems with
vacua having negative cosmological constant \cite{BFLR4}, which tend to
dominate the probabilities of observations and hence would make our
observation of a positive cosmological constant highly improbable.  For
the scale-factor and fat-geodesic measures, this is because the
geodesics can go for a very long proper time in a vacuum with a very
tiny negative cosmological constant before reaching the scale-factor
cutoff or the big crunch.  For the causal-patch measure, it is because
the causal patch can be very large in a region with a very small
negative cosmological constant.  This is essentially the same problem
that arises with those measures for the Minkowski vacuum, except that
here one gets the domination by ordinary observers in excitations of
vacua with negative cosmological constants rather than by Boltzmann
brains in the Minkowski vacuum.

On the other hand, since the Agnesi weighting damps late times whether
or not geodesics are exponentially diverging or whether or not a causal
patch has a bounded spatial size, it suppresses the late-time
contributions of not only the Minkowski vacuum but also all vacua with
positive or negative cosmological constant.  So long as the quantum
state does not strongly favor negative values of the cosmological
constant, there is nothing in Agnesi weighting that would favor them
either, so there is no statistical conflict with our observations of a
positive cosmological constant. 

\baselineskip 17 pt

\section*{The youngness effect}

The combination of spatial density averaging (volume averaging) and
Agnesi weighting does lead to a very mild youngness effect, because the
expectation value $dN_j/dt$ of the number of observations $O_j$ per
proper time is divided by both the 3-volume $V(t)$ of the hypersurface
and by the Agnesi factor $1+t^2$.  This tends to favor observations
early in the universe, so let us see how great a youngness effect it
gives, say between the origin of the solar system and a time equally far
in the future, near its expected demise.

Let us use what I call the Mnemonic Universe Model (MUM, which itself
might be considered a British term of endearment for Mother Nature) for
the universe, a spatially flat universe dominated by dust and a
cosmological constant, with present age $t_0 = H_0^{-1} = 10^8\
\mathrm{years}/\alpha$, where $\alpha \approx 1/137036000$ \cite{PPB} is
the electromagnetic fine structure constant, and with the solar age
$t_0/3$.  The present observations give a universe age of $13.69\pm
0.13$ Gyr \cite{PPB} that is $0.999\pm 0.009$ times the MUM value of
13.7036 Gyr, a Hubble constant of $72\pm 3$ km s$^{-1}$ Mpc$^{-1}$
\cite{PPB} that is $1.009\pm 0.042$ times the MUM value of 71.3517 km
s$^{-1}$ Mpc$^{-1}$, and a solar system age of $4.5681\pm 0.0003$ Gyr
\cite{BW} that is $1.00005\pm 0.00007$ times the MUM value of 4.56787
Gyr.  Thus the MUM values are all within the present observational
uncertainties for the universe age, Hubble constant, and solar system
age.

The metric for the MUM model is
\begin{equation} 
ds^2 = -dt^2 + \sinh^{4/3}(1.5H_\Lambda t)(dx^2+dy^2+dz^2),
\label{MUM} 
\end{equation} 
where $H_\Lambda = \sqrt{\Lambda/3}$ is the asymptotic value of the
Hubble expansion rate
\begin{equation} 
H = \frac{\dot{a}}{a} = H_\Lambda\coth{(1.5H_\Lambda t)}.
\label{H} 
\end{equation} 

For $t_0 = H_0^{-1}$, we need $H_\Lambda t_0 = \tanh{(1.5H_\Lambda
t_0)}$ or $H_\Lambda t_0 \approx 0.858560$, and then $t_0 = 10^8
\mathrm{years}/\alpha$ gives $H_\Lambda \approx (15.96115
\mathrm{Gyr})^{-1} \approx 61.2597$ km s$^{-1}$ Mpc$^{-1}$.  One can
also calculate that the MUM predicts that at present the dark energy
corresponding to the cosmological constant gives a fraction of the total
(closure) energy density that is $\Omega_\Lambda = \tanh^2{(1.5H_\Lambda
t_0 )} = (H_\Lambda t_0)^2 \approx 0.737125$, in good agreement with the
observational value of $0.74\pm 0.03$ \cite{PPB} that is $1.004 \pm
0.041$ times the MUM value.

Some features of the MUM are that with the conformal time that is given
by $\eta = \int_0^t dt'/\sinh^{2/3}(1.5H_\Lambda t')$, the total
conformal time is $\eta_\infty \approx 44.76088$ Gyr, and the present
value of the conformal time is $\eta_0 \approx 33.8825\ \mathrm{Gyr}
\approx 0.756967\eta_\infty$.  (This is using the normalization above
that $a(t) = \sinh^{2/3}(1.5H_\Lambda t)$, which gives $a_0 \equiv
a(t_0) \approx 1.41014$; if one had instead set $a_0 = 1$ so $a(t) =
\sinh^{2/3}(1.5H_\Lambda t)/\sinh^{2/3}(1.5H_\Lambda t_0)$, one would
have $\eta = \int_0^t dt'/a(t')$ giving $\eta_\infty \approx 63.1193$
Gyr and $\eta_0 \approx 47.7792$ Gyr.)  Thus we see that although there
is only a finite proper time in the past and an infinite proper time in
the future, over three-quarters of the total finite conformal time of
the MUM has already passed.

The cosmological event horizon for the comoving observer at $r =
\sqrt{x^2+y^2+z^2} = 0$ (which we shall take to be our worldline) is at
$r = \eta_\infty - \eta$, so on the constant-time hypersurface $t=t_0$
(and hence $\eta = \eta_0$), it is at $r = r_1 = \eta_\infty - \eta_0
\approx 10.8784$ Gyr, at a distance along this hypersurface of $a_0 r_1
\approx 15.3401$ Gyr (times the speed of light $c$, which I am setting
to unity; e.g., this distance is 15.3401 billion light years).  The
actual spacetime geodesic distance from us to the point on the comoving
worldline at $r = r_1$ that is crossing our cosmological event horizon
when its proper time from the big bang is the same as ours is 16.2282
Gyr, greater than the distance along a geodesic of the constant-time
hypersurface, because geodesics of that hypersurface are not geodesics
of spacetime but instead bend in the timelike direction, shortening
their length.  The actual geodesic of spacetime joining the two events
goes forward in the time $t$ from $t_0$ to $t \approx 1.17686 t_0
\approx 16.1272$ Gyr, to a point with $a \approx 1.18725 a_0$, before
bending back in $t$ to get back to $t_0$ at the cosmological event
horizon.

Like de Sitter spacetime with the same value of the cosmological
constant, the MUM has a maximal separation of two events connected by a
spatial geodesic, which is $\pi/H_\Lambda \approx 50.1434$ Gyr.  All
events with $r \geq 2\eta_\infty - \eta_0 - \eta$ cannot be reached by
any geodesics from our location in spacetime.  The events on this
boundary at $t=t_0$ are at $r = r_2 = 2(\eta_\infty - \eta_0) \approx
21.7567$ Gyr, which is at a distance of 30.6802 Gyr along the $t=t_0$
hypersurface, though the geodesic distance is the maximal value of
50.1434 Gyr.  (Actually, there is no geodesic to this boundary itself,
but this maximal value is the limit of the geodesic distance as $r$
approaches the boundary.)

A third preferred distance on the $t=t_0$ hypersurface of homogeneity is
at $r = r_3 = \eta_0 \approx 33.8825$ Gyr, which is where a comoving
worldline that started at the big bang on our past light cone reaches
after the same proper time $t_0$ from the big bang as we are.  That is,
this is the present location of a worldline which started at our
particle horizon.  This value of $r$ corresponds to a physical distance
along this hypersurface of 47.7792 Gyr.  There are no geodesics from us
to that point, so even if we had a tachyon gun, we could not hit that
worldline at a point on it after its proper time passed our value of
$t_0$.

The MUM also allows on to calculate the geodesic distance from us to
each of these three worldlines along a geodesic that is orthogonal to
our worldline at its intersection here and now.  This distance to $r =
r_1$ (the comoving worldline that crosses our cosmological event horizon
at a proper time of $t_0$) is 11.3244 Gyr, to $r = r_2$ (the worldline
that after proper time $t_0$ reaches the boundary of where geodesics
from us can reach) is 14.3274 Gyr, and to $r = r_3$ (the worldline that
starts at the big bang on our past light cone) is 14.6863 Gyr.  This
spacelike geodesic never reaches our cosmological event horizon but
instead ends at the big bang at a distance of 14.6889 Gyr from us, where
$r = r_4 \approx 41.0459$ Gyr (or $a_0 r_4 \approx 57.8806$ Gyr for the
distance along the $t = t_0$ hypersurface to the comoving worldline with
$r = r_4$), which is less than the value $r = r_5 \approx 44.7609$ Gyr
where our cosmological event horizon intersects the big bang, whose
comoving worldline is at a distance $a_0 r_5 \approx 63.1193$ Gyr from
us along the $t = t_0$ hypersurface.  That is, if we define simultaneity
by spacelike geodesics orthogonal to our worldline, the big bang is
still going on right now \cite{howbig}, at a distance of 14.7 billion
light years from us in the Mnemonic Universe Model.

Yet another comoving worldline that one may define is the one that
emerges from the big bang from the boundary of the region that can be
reached from us by spacetime geodesics.  This is at $r = r_6 =
2\eta_\infty - \eta_0 \approx 55.6393$ Gyr, which as measured along the
$t=t_0$ hypersurface is at the distance $a_0 r_6 \approx 78.4594$
billion light years from us.  This is the upper limit to the current
distance (over a constant-time hypersurface, not along a spatial
geodesic of spacetime that has a maximum length of 50.1434 billion light
years in the MUM) of any comoving worldline that can be reached by any
geodesics from our current location in spacetime. The limit of the
spatial geodesics that reach from us to comoving worldlines as $r
\rightarrow r_6$ is a null geodesic that goes from us to the spacelike
future boundary at $\eta = \eta_\infty$ and then returns to the big bang
along another null geodesic; the spacelike geodesics approaching this
limit approach the maximum spacelike geodesic length of 50.1434 billion
light years, this length occurring in the de Sitter region in the
arbitrarily distant future where the spatial geodesic turns around from
going toward the future in $t$ to going back toward the past in $t$.

Now let us use the MUM to calculate the youngness effect from the
formation of the solar system, at a time $t_0/3$ before the present, or
at $t=2t_0/3$ after the big bang, to a time equally equally far in the
future, at $t=4t_0/3$, which we shall use as a very crude approximation
for the mnemonic demise of the solar system.  Since both of these times
are enormously longer than the Planck time (with $t_0 = 8.021\times
10^{60}$ in Planck units), we can drop the 1 that is included in the
Agnesi weighting to avoid a divergence at $t=0$.  Then we see that on a
per-time basis, the Agnesi weighting factor of $1/(1+t^2)$ is four times
smaller at the demise of the solar system than at its formation. 
However, the spatial volume of the universe also goes up by a factor of
7.75 during this `lifetime' of the solar system, so if we had a fixed
comoving density of observers, the combination of the Agnesi and spatial
density averaging (volume averaging) factors would give about 31 times
the weight for observations at the formation of the solar system than at
its end.

This would imply that if the same number of observations occurred per
proper time and per comoving volume throughout the lifetime of the solar
system, the ones at the demise would have only about 3\% of the measure
of the ones at the formation.  Half of the measure would occur within
the first 18\% of the solar system lifetime.  This effect would tend to
favor observations early in the history of the solar system.  

However, it seems highly plausible that a factor of only about 31 would
be negligible in comparison with the factors that determine the numbers
of observations.  Presumably if one sampled a huge number of solar
systems, only a very tiny fraction of the observations would occur very
close to the formation, because of the time needed for evolution. If the
probability for evolution to intelligent life to have occurred rises
sufficiently rapidly with the time after the formation time (e.g.,
significantly faster than the linear rise one would expect if evolution
were a single event that occurred statistically at a constant rate per
time per solar system), then it would not be at all surprising that we
exist at a time when 85\% of the measure would have passed {\it if} the
number of observations were instead uniform in time.

The shift of the measure (say calibrated for a fixed comoving density of
observers making a constant number of observations per time) from being
uniform in the time to having the weighting factors of the inverse
three-volume (from the spatial density averaging) and of very nearly the
inverse square of the time (from the Agnesi weighting) would have an
effect on the number of hard steps $n$ Brandon Carter estimated for the
evolution of intelligent life on earth \cite{Carter83,Carter07}.  A hard
step (or `critical' step in the first of these papers) is one whose
corresponding timescale is at least a significant fraction of the
available time for it to occur (e.g., the lifetime of the sun).  Carter
emphasized \cite{Carter83} that unless there is an unexplained (and
therefore {\it a priori} improbable) coincidence, the timescale of a
step is not likely to be close to the available time, so generically a
hard step has a timescale much longer than the available time. 
Therefore, a hard step is unlikely to occur within the available time on
a random suitable planet in which the previous steps have occurred.

In the first of these papers \cite{Carter83}, Carter assumed that since
we are about halfway through the predicted lifetime of the sun, we arose
about halfway through the life-permitting period on earth and about
halfway through the measure if the measure were uniform in time.  He
then concluded that the number of hard steps $n$ would likely be 1 or
2.  In the second paper \cite{Carter07}, Carter used more recent
information \cite{Caldiera-Keating} that the sun may become too luminous
for life to continue on earth just one billion years in the future
rather than five.  Then we would be a fraction $f \sim 5/6$ of the way
through the available period for life, and this would lead to an
estimate for the number of hard steps to be $n \sim f/(1-f) \sim 5$. 
(Carter suggested $4 \stackrel{<}{\sim} n \stackrel{<}{\sim} 8$ and
favored $n=6$ if the first hard step occurred on Mars.)

Now let us see how these estimates for the number of hard steps to us
would be modified with the spatial density averaging and Agnesi
weighting.  If we take the assumptions of Carter's original paper, that
the available time is the entire solar lifetime and that we are halfway
through it, without any measure factors $f$ would be $0.5$, but with my
measure factors this fraction would be changed to $f = 0.85$, which
would then give $n \sim 6$ even without the natural global warming
effects of rising solar flux.  On the alternative assumption that there
is only one gigayear left for life on earth, my measure factors change
Carter's $f = 5/6$ to $f = 0.94$ and hence give the number of hard steps
as $n \sim f/(1-f) \sim 16$.

Therefore, if we could really learn what the number of hard steps were
for the evolution of intelligent life here on earth, we could in
principle test between different proposals for the measure, such as
between the scale-factor measure and my proposed spatial density
averaging with Agnesi weighting.  However, this currently seems like a
very hard problem.  (Would it be another hard step for intelligent life
to solve it?)  All I can say at present is that it does not seem
obviously in contradiction with observations that the number of hard
steps might be higher than Carter's estimates, so our present knowledge
does not appear to provide strong evidence against the proposed spatial
density averaging and Agnesi weighting. 

\baselineskip 19 pt

\section*{Conclusions}

Agnesi weighting gives a precise weighting factor that may be an
improvement over the indefinite cutoff proposed by other proposals, such
as the scale-factor measure \cite{SGSV,BFYb,DGLNSV,SiSa}.  Unlike what
occurs in the latter, in which time comes to a sharp end at an
unspecified time \cite{BFLR2}, in Agnesi weighting old universes never
die, they just fade away.  This fading away is purely in the measure for
the various observations and not in any property of the contents of the
observations themselves (e.g., of the observed spectrum of the CMB, or
of how painful a toothache feels), so it cannot be directly observed. 
However, if one did have an observation that a theory with this fading
said would have excessively low measure, that would be statistical
evidence against that theory. One can make similar statistical
interpretations of the idea that time ends abruptly at an unspecified
time, so that difference by itself is a matter of the assumed ontology
rather than of different testable statistical predictions.

When combined with number density averaging (which I previously called
volume averaging \cite{Born1,Born2,Born3,Born4,Born5}) and with a
suitable quantum state for the universe (such as the Symmetric-Bounce
state \cite{Symbounce}), Agnesi weighting gives a finite measure for
observations in the universe and appears to avoid the Boltzmann brain
problem and other potential problems of cosmological measures, even
without restrictions on the decay rates of anthropic vacua used to solve
the Boltzmann brain problem in other measures
\cite{Linde06,BF,BFYb,DGLNSV}, even allowing for Boltzmann brains to
form in the 11-dimensional Minkowski vacuum that states in the string
landscape may lead to \cite{Page05,Page06b,Page06c,Page06d,Hartle}, and
even allowing vacua with negative cosmological constants that tend to
dominate the probabilities in other measures \cite{BFLR4}.  Agnesi
weighting leads to a very mild youngness effect, but one which is well
within the current uncertainties of how rapidly intelligent life is
likely to have evolved on earth.

Phenomenologically, Agnesi weighting appears to work well.  However, it
is surely not the last word on the subject.  For one thing, although it
is quite simple, it is rather {\it ad hoc} (like all other solutions to
the measure problem proposed so far, at least in my mind), so one would
like to learn some principle that would justify it or an improvement to
it.  Second, it is presently formulated only in the semiclassical
approximation to quantum cosmology, so one would want a fully quantum
version.  These challenges will be left for future work.

\section*{Acknowledgments}

The idea for this paper came while I was visiting James Hartle, Donald
Marolf, Mark Srednicki, and others at the University of California at
Santa Barbara in February 2010.  I am especially grateful for the
hospitality of the Mitchell family and of the George P. and Cynthia W.
Mitchell Institute for Fundamental Physics and Astronomy of Texas A\&M
University at a workshop in April 2010 at Cook's Branch Conservancy,
where I had many more valuable discussions on this subject, especially
with James Hartle, Stephen Hawking, and Thomas Hertog. I was further
stimulated by discussions with many colleagues at Peyresq Physics 15 in
June 2010 under the hospitality of Edgard Gunzig and OLAM, Association
pour la Recherche Fondamentale, Bruxelles.  Comments by Raphael Bousso,
both by email and while I was enjoying his hospitality at the University
of California at Berkeley in February 2011, were instrumental for
various revisions of the paper, as well as further discussions with
James Hartle, Stefan Leichenauer, Vladimir Rosenhaus, and Edward
Witten.  This research was supported in part by the Natural Sciences and
Engineering Research Council of Canada.

\end{document}